\documentclass[aps,prb,amsmath,amssymb,twocolumn,superscriptaddress]{revtex4}
\usepackage{graphicx}
\usepackage{braket}
\usepackage{xcolor}
\usepackage{hyperref}
\usepackage[normalem]{ulem}
\usepackage{soul}

\begin{document}


\title{Simultaneous operations in a two-dimensional array of singlet-triplet qubits}

\author{Federico~Fedele} 
\thanks{These authors contributed equally to this work}
\affiliation{Center for Quantum Devices, Niels Bohr Institute, University of Copenhagen, 2100 Copenhagen, Denmark}

\author{Anasua~Chatterjee}
\thanks{These authors contributed equally to this work}
\affiliation{Center for Quantum Devices, Niels Bohr Institute, University of Copenhagen, 2100 Copenhagen, Denmark}

\author{Saeed~Fallahi}
\affiliation{Department of Physics and Astronomy, Purdue University, West Lafayette, Indiana 47907, USA}
\affiliation{Birck Nanotechnology Center, Purdue University, West Lafayette, Indiana 47907, USA}

\author{Geoffrey~C.~Gardner}
\affiliation{Microsoft Quantum Lab Purdue, Purdue University, West Lafayette, Indiana 47907, USA}
\affiliation{Birck Nanotechnology Center, Purdue University, West Lafayette, Indiana 47907, USA}

\author{Michael~J.~Manfra}
\affiliation{Department of Physics and Astronomy, Purdue University, West Lafayette, Indiana 47907, USA}
\affiliation{Schools of Materials Engineering and Electrical and Computer Engineering, Purdue University, West Lafayette, Indiana 47907, USA}
\affiliation{Birck Nanotechnology Center, Purdue University, West Lafayette, Indiana 47907, USA}
\affiliation{Microsoft Quantum Lab Purdue, Purdue University, West Lafayette, Indiana 47907, USA}

\author{Ferdinand~Kuemmeth}
\affiliation{Center for Quantum Devices, Niels Bohr Institute, University of Copenhagen, 2100 Copenhagen, Denmark}

\newcommand{\VL}{V_\mathrm{L}}
\newcommand{\VM}{V_\mathrm{M}}
\newcommand{\VR}{V_\mathrm{R}}

\newcommand{\Btot}{B^\mathrm{tot}}
\newcommand{\Bext}{B^\mathrm{ext}}
\newcommand{\Bznuc}{B_\mathrm{z}^\mathrm{nuc}}
\newcommand{\Bpnuc}{B_\perp^\mathrm{nuc}}

\newcommand{\ud}{\uparrow\downarrow}
\newcommand{\du}{\downarrow\uparrow}
\newcommand{\drv}{\mathrm{d}}

\newcommand{\FHahn}{F_\mathrm{Hahn}}
\newcommand{\FCPMG}{F_\mathrm{CPMG}}
\newcommand{\FFID}{F_\mathrm{FID}}
\newcommand{\Fenv}{F_\mathrm{env}}

\newcommand{\Ga}{^{69}\mathrm{Ga}}
\newcommand{\Gb}{^{71}\mathrm{Ga}}
\newcommand{\As}{^{75}\mathrm{As}}
\newcommand{\fGa}{f_{^{69}{\rm Ga}}}
\newcommand{\fGb}{f_{^{71}{\rm Ga}}}
\newcommand{\fAs}{f_{^{75}{\rm As}}}

\newcommand{\TCPMG}{T_2 ^\mathrm{CPMG}}
\renewcommand{\vec}[1]{{\bf #1}}

\newcommand{\remove}[1]{{\color{red} \sout{#1}}}
\newcommand{\add}[1]{{\color{black} #1}}
\newcommand{\com}[1]{\emph{\color{cyan} #1}}
\definecolor{darkgreen}{rgb}{0,0.6,0}
\newcommand{\FF}[1]{\emph{\color{darkgreen} #1}}
\newcommand{\FK}[1]{\emph{\color{magenta} #1}}
\newcommand{\lookat}[1]{{\color{orange} #1}}
\date{\today}

\maketitle
\textbf{
In many physical approaches to quantum computation, error-correction schemes assume the ability to form two-dimensional qubit arrays with nearest-neighbor couplings and parallel operations at multiple qubit sites. 
While semiconductor spin qubits exhibit long coherence times relative to their operation speed and single-qubit fidelities above error correction thresholds, multi-qubit operations in two-dimensional arrays have been limited by fabrication, operation, and readout challenges. We present a two-by-two array of four singlet-triplet qubits in gallium-arsenide and show simultaneous coherent operations and four-qubit measurements via exchange oscillations and frequency-multiplexed single-shot measurements. A larger multielectron quantum dot is fabricated in the center of the array as a tunable inter-qubit link, which we utilize to demonstrate coherent spin exchange with \add{selected qubits}. Our techniques are extensible to other materials, indicating a path towards quantum processors with gate-controlled spin qubits.}

\section{Introduction}
Semiconducting spin qubits are one of the leading candidates for enabling quantum computation and have demonstrated relatively long coherence times~\cite{Watson2017,Muhonen2014}, figures of merit at the fault tolerance threshold~\cite{Yoneda2018}, and high-fidelity single and two-qubit gates~\cite{He2019,Watson2018,Zajac2018,Veldhorst2015}. 
Various materials are being investigated, including gallium-arsenide (GaAs), silicon, and germanium structures~\cite{Chatterjee2021}. 
While linear and two-dimensional arrays have been reported up to 9$\times$1 and 3$\times$3 qubit sites~\cite{Zajac2016,Mortemousque2021}, respectively, most state-of-art GaAs~\cite{kojima2020,Ito2018,Qiao2020}, silicon~\cite{Watson2017,Sigillito2019,takeda2020} and germanium~\cite{Hendrickx2021} experiments only performed individual single- and two-qubit operations at a time. 
(For specific applications in GaAs, however, simultaneous interdot exchange couplings have been demonstrated~\cite{Hensgens2017, Malinowski2018,Dehollain2020,Qiao2020}.) 
Obstacles for scaling gate-controlled spin qubit operations to larger arrays include the placement of proximal charge sensors for multi-qubit readout, crosstalk between gate electrodes, and reserving space for gate fan-out while preserving nearest-neighbour qubit couplings.

In this work we simultaneously operate and read out four singlet-triplet ($S$-$T_{0}$) qubits, arranged around a central quantum dot in a two-by-two array with four integrated charge sensors. Synchronized voltage pulses applied to eight gate electrodes initialize all qubits and induce coherent Overhauser or exchange rotations.    
Each four-qubit rotation is followed by spin-to-charge conversion of all qubits, by reading out four local charge sensors via frequency-multiplexed reflectometry. 
Single-shot measurement outcomes as a function of exchange (dephasing) time clearly reveal simultaneous exchange oscillations (Overhauser rotations) in all four qubits, with quality factors ranging between 3.4 and 5.1, and inhomogeneous dephasing times around 20 ns. 
By interlacing dephasing and exchange operations, we observe a correlation between low Overhauser gradients and low visibilities of exchange oscillations. 
\add{
The speed of exchange rotations is used to detect a small crosstalk between qubits. 
By applying voltage pulses to the central multielectron dot, intended as a coupler, we also demonstrate coherent qubit-coupler spin exchange processes. (Qubit-qubit spin exchange, mediated by the coupler, is not demonstrated in this work.) 
}
The size of the coupler creates sufficient spacing between qubits to facilitate fan-out of gate electrodes and addressability of qubits for individual operation and readout, while still retaining a small overall footprint that makes spin qubits attractive for scaling to large systems. 

These techniques constitute a tool set for building  quantum processors with programmable connectivities between spin qubits, and may find use also in silicon and germanium quantum circuits.    

\begin{figure*}
	\includegraphics[scale=1]{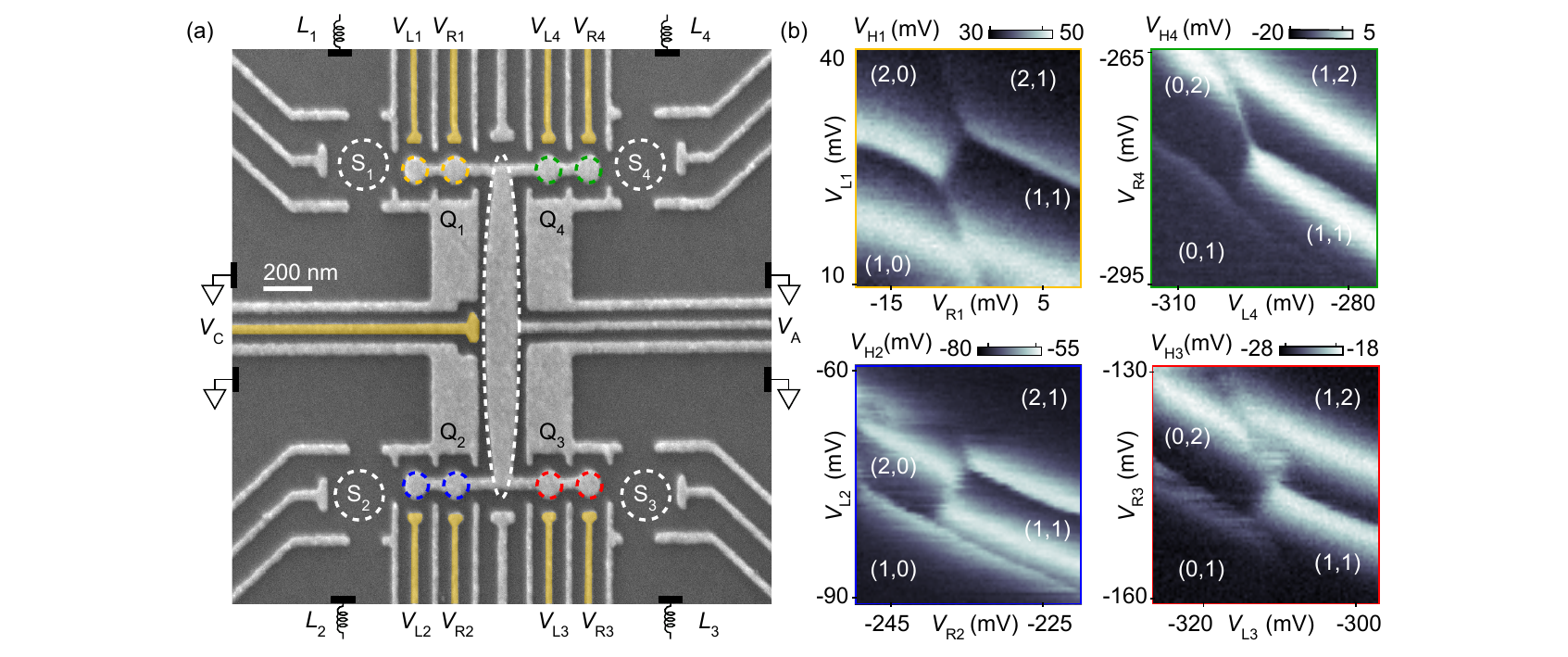}
	\caption{Two-dimensional array of four tunable GaAs double dots, each implementing one singlet-triplet qubit. (a) Scanning electron micrograph of a device similar to the one measured. 
All gate electrodes (light gray) are connected to independent tuning voltages ($V_i$). 
Electron reservoirs (black bars) are connected to ohmics, four of which are wirebonded to inductors $L_{1-4}$. 
Four double dots (small dashed circles) coupled to an elongated multielectron quantum dot (dashed ellipse) are formed by applying a positive voltage to the accumulation gate ($V_\mathrm{A}$). 
The charge state of each qubit is detected via proximal sensor dots (S$_\mathrm{1-4}$) by applying frequency-multiplexed reflectometry to $L_{1-4}$ and recording the associated homodyne-detected voltages $V_\mathrm{H1-H4}$. 
Gate electrodes false-colored in yellow are connected via bias tees to coaxial lines, allowing the application of fast manipulation pulses $V_\mathrm{C}(t)$, $V_{\mathrm{L}i}(t)$, $V_{\mathrm{R}i}(t)$ in Figures 2-4. 
(b) Charge stability diagrams of the four singlet-triplet qubits in the absence of pulses. Numbers in parentheses indicate the charge configuration of each double dot. } 	
	\label{fig1}
\end{figure*}

\section{Results}
All measurements, except where indicated, were performed in a dilution refrigerator at a base temperature of 25 mK in the presence of a static magnetic field of $B_{\parallel}$=120~mT applied in the plane of the sample.

\subsection{Device, multiplexed setup and tuning}

Figure~\ref{fig1}a shows the two-dimensional layout of the quantum dot circuit, with the four qubits arranged in a $2$x$2$ array (Q$_{1}$,  Q$_{2}$, Q$_{3}$ and Q$_{4}$). 
Each qubit is encoded in a double quantum dot (DQD) and operated as a singlet-triplet spin qubit~\cite{Petta2005}. 
Sensor quantum dots (S$_\mathrm{1-4}$) are placed on the outside of each DQD for single-shot qubit readout via spin-to-charge conversion~\cite{Barthel2010}. 
For each sensor, this geometry gives good contrast between the charge states within its DQD, while four depletion gates with larger (rectangular) area help screening the electrostatic crosstalk from the other three DQDs. 

The gate design also features an elongated gate at the center of the array, connected to circular regions under which the DQDs are formed. This gate is operated in accumulation mode, which not only improves dot confinements and gate control~\cite{Martins2016}, but also accumulates a multielectron dot in the large elongated potential well (white dashed line). This central dot can be pulsed by control voltage $V_\mathrm{C}$ (intended to mediate coherent spin-exchange between any two qubits) and can also serve as an inner electron reservoir for the DQDs. Fast voltage pulses can also be applied to gates labeled $V_{\mathrm{R}i,\mathrm{L}i}$ (false-colored in gold in Fig.~\ref{fig1}a), allowing fast manipulation of all DQDs.   
All pulsed gate electrodes are wirebonded to bias-tees located on a high-bandwidth PCB sample holder where low-frequency tuning voltages and high-frequency control pulses are combined~\cite{Qdevil}. On the daughter board of the sample holder, four inductors with different inductances ($L_1$=1200 nH, \add{$L_2$=560} nH, $L_3$= 750 nH, \add{$L_4$= 910 nH}) are capacitively coupled to one reflectometry channel of the cryostat, allowing frequency-multiplexed readout of all sensor dots~\cite{Laird2010}. 

Figure~\ref{fig1}b shows charge stability diagrams for all DQDs, measured by digitizing the demodulated reflectometry signal $V_{\mathrm{H}i}$ of the proximal sensor dots S$_{i}$ (see below) as a function of the respective plunger gate voltages $V_{\mathrm{R}i}$ and $V_{\mathrm{L}i}$. 
All devices are tuned to the (1,1) charge state with the (2,0)-(1,1) tunnel couplings adjusted to allow coherent operations as singlet-triplet qubits. 
(Numbers in parenthesis ($l$,$r$) represent the number of electrons on the left and right quantum dot, respectively.)  

\begin{figure*}
	\includegraphics[scale=1.1]{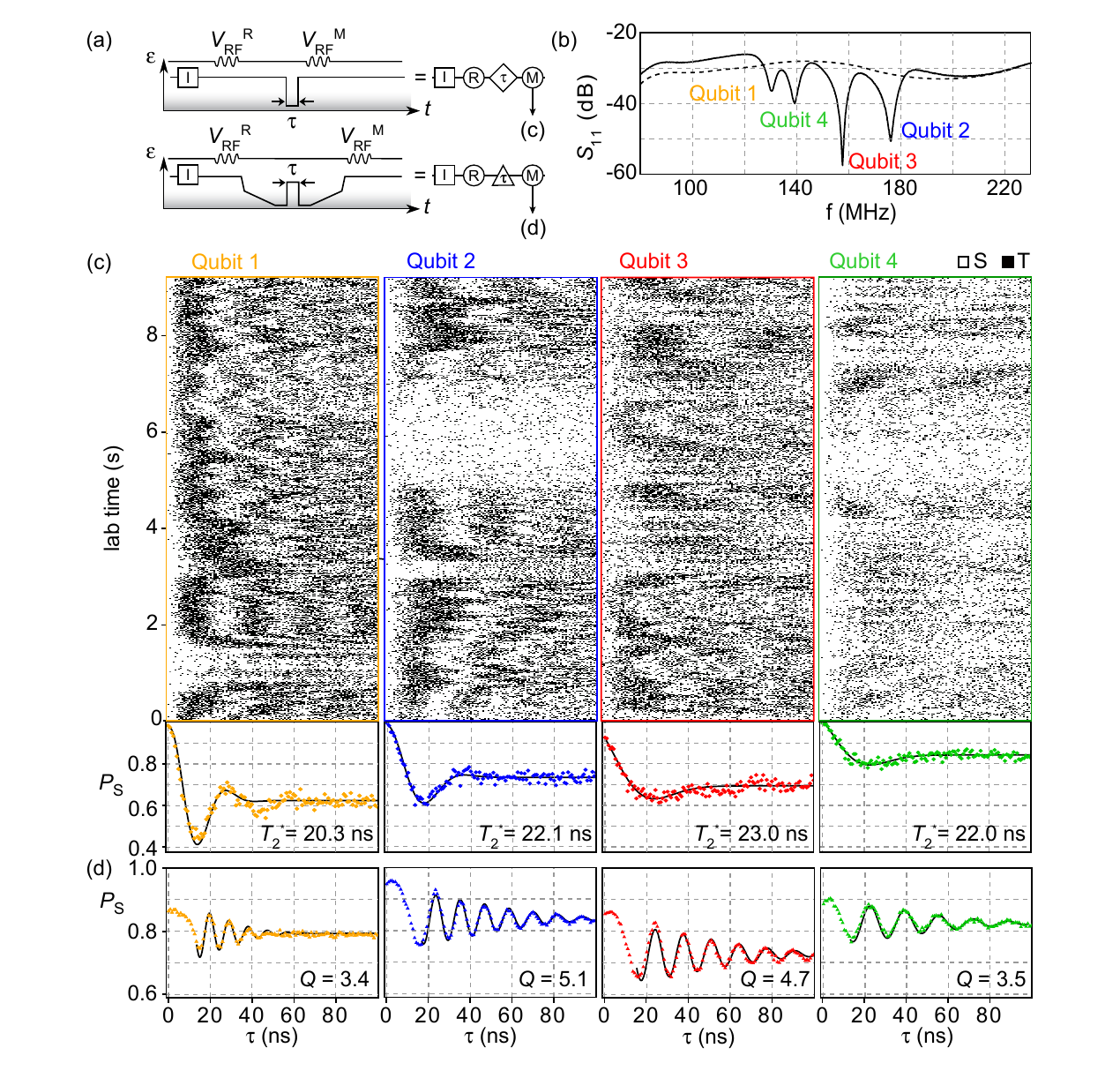}
	\caption{Simultaneous four-qubit operations. (a) Detuning cycle $\varepsilon (t)$, applied simultaneously to all qubits by appropriate voltage pulses $V_{\mathrm{L}i}(t)$, $V_{\mathrm{R}i}(t)$, resulting in dephasing operations (top) or exchange rotations (bottom). Qubit initialization (I) is followed by a reference measurement (R) of the reflectometry signal ($V_\mathrm{RF}^\mathrm{R}$), a dephasing or exchange pulse of duration $\tau$, and a measurement (M) of the reflectometry signal ($V_\mathrm{RF}^\mathrm{M}$) immediately after the final rise of $\varepsilon (t)$ (spin-to-charge conversion). 
(b) Reflection coefficient of the reflectometry setup as a function of the carrier frequency, in the presence (black line) and absence (dashed line) of tuning voltages applied to all gate electrodes. Four distinct resonances, arising from different inductance values used for $L_{1-4}$, allow frequency-multiplexed single-shot readout of all four qubits. 
(c) Single-shot outcomes of 61440 dephasing cycles applied simultaneously to all four qubits. Each singlet outcome is represented by one white pixel. Each row represents 120 cycles with increasing values for $\tau$. Acquisition time for 512 rows is 9 seconds. Bottom panels show the fraction of singlet outcomes, $P_\mathrm{S}$, obtained by averaging the top panels along the laboratory time. Fits to the data (solid lines) yield $T_{2}^{*}$  times similar for all four qubits.
(d) Fraction of singlet outcomes, $P_{\mathrm{S}}(\tau)$, for 61440 exchange-rotation cycles applied simultaneously to all four qubits.}
	\label{fig2}
\end{figure*}

\subsection{Simultaneous measurements}
The distinct resonance frequencies associated with the inductors allow simultaneous measurements within the array, either by continuous excitation of all inductors (as was done in Fig.~\ref{fig1}b), or by application of short readout bursts within each qubit operation cycle. 
Each qubit cycle involves a detuning pulse applied to all qubits, taking each DQD from (2,0) to (1,1) and back to (2,0) as illustrated in Fig.~\ref{fig2}a as a decrease and increase of the detuning parameter $\varepsilon(t)$. For each qubit, comparison of the demodulated reflectometry voltage during the measurement burst ($V_\mathrm{RF}^\mathrm{M}$) with the demodulated voltage during the reference burst ($V_\mathrm{RF}^\mathrm{R}$, right after initialization of the DQD) allows assignment of single-shot readouts to each qubit, namely singlet if the measurement burst indicates a (2,0) charge state or triplet if the measurement burst indicates a (1,1) charge state.  

Each reference and measurement burst applied to the cryostat comprises the four resonance frequencies, yielding qubit-resolved demodulated voltages $V_{\mathrm{H}i}$ due to the large separation of resonances in frequency domain (see Methods). 
Figure~\ref{fig2}b shows the reflection coefficient of the sample holder~\cite{Reilly2007}. 
Four well-separated resonances between $130-180$ MHz are clearly visible, which disappear one by one if gate voltages associated with the sensor dots are deactivated (dashed line)~\cite{Laird2010}. 
(The overall low $S_{11}$ values are caused by intentional attenuation within the cryostat, which exceeds the gain of the cryogenic amplifier in the reflectometry channel.)  
During an experiment, the reflected bursts are amplified and then demodulated by four parallel homodyne mixers. 
The resulting four demodulated voltages $V_{\mathrm{H}i}$ are measured by a four-channel digitizer, and each channel analyzed to assign single-shot outcomes (Methods). 
The optimal burst duration depends on the qubit relaxation time and sensor signal-to-noise ratio. 
With a typical integration time of 10 $\mu$s, we can distinguish (1,1) from (2,0) outcomes with a signal-to-noise ratio of approximately 5.

Figure~\ref{fig2}c demonstrates simultaneous four-qubit single-shot readout after each qubit has precessed freely in the Overhauser gradient within its DQD. 
Each pixel, white for singlet outcome and black for triplet outcome, results from application of one 8-dimensional gate-voltage cycle that implements the top cycle in Fig.~\ref{fig2}a for each qubit by pulsing $V_{\mathrm{L}i}$ and $V_{\mathrm{R}i}$~\cite{Malinowski2017}. 
After the qubit is initialized in its (2,0) singlet state (I), a detuning pulse quickly separates the two electrons by pulsing towards (1,1) where spin procession between $\ket{S}$ and $\ket{T_{0}}$ is driven by the Overhauser field gradient. After precession time $\tau$, the DQD is pulsed back towards (2,0) to implement spin-to-charge conversion via Pauli blockade. 
Repeating this pulse cycle for different values of $\tau$ yields one row for all qubits. 
A plot of 512 rows obtained right after each other clearly shows four different Overhauser gradients fluctuating over a period of 9 seconds at four different locations of the same chip. 
While no obvious qubit-qubit correlations are visible in this data, an extension of the systematic study in Ref. \onlinecite{Malinowski2017} to four qubits or an implementation of dynamical noise spectroscopy pulses that are specifically designed to measure spatio-temporal qubit-qubit correlations~\cite{Luke2016} may provide useful insights for optimizing high-fidelity multi-qubit operations. 

To estimate the inhomogeneous dephasing time $T_{2}^{*}$ for each qubit, we plot at the bottom of Fig.~\ref{fig2}c the fraction of singlet outcomes, $P_\mathrm{S}$, for each value of $\tau$. 
Fitting a semiclassical model of dephasing~\cite{Petta2005} to the resulting decays yields $T_{2}^{*}$ times of approximately 20~ns, which is typical for singlet-triplet qubits in GaAs. 

Figure~\ref{fig2}d shows coherent exchange oscillations induced simultaneously in all qubits, by plotting $P_\mathrm{S}$ for the lower pulse cycle in panel~\ref{fig2}a. Here, an adiabatic detuning ramp from (2,0) to (1,1) prepares each qubit in the eigenstate of the Overhauser gradient before a quick detuning pulse of duration $\tau$ is applied that turns on an exchange interaction between the two spins. After the exchange pulse, an adiabatic ramp converts the resulting state into either singlet (2,0) or triplet (1,1) charge states. Coherent exchange oscillations in $P_\mathrm{S}(\tau)$ are clearly observed, with quality factors ($3.4-5.1$) similar to those observed for non-symmetric exchange pulses in single-qubit devices~\cite{Martins2016}. 
The frequency of exchange oscillations can be tuned individually for each qubit by changing the amplitude of the detuning pulse, and has been chosen for Figure~\ref{fig2}d to yield a
 $\pi$ rotation in 15 ns for all qubits. 
 

\begin{figure}
	\includegraphics[scale=1]{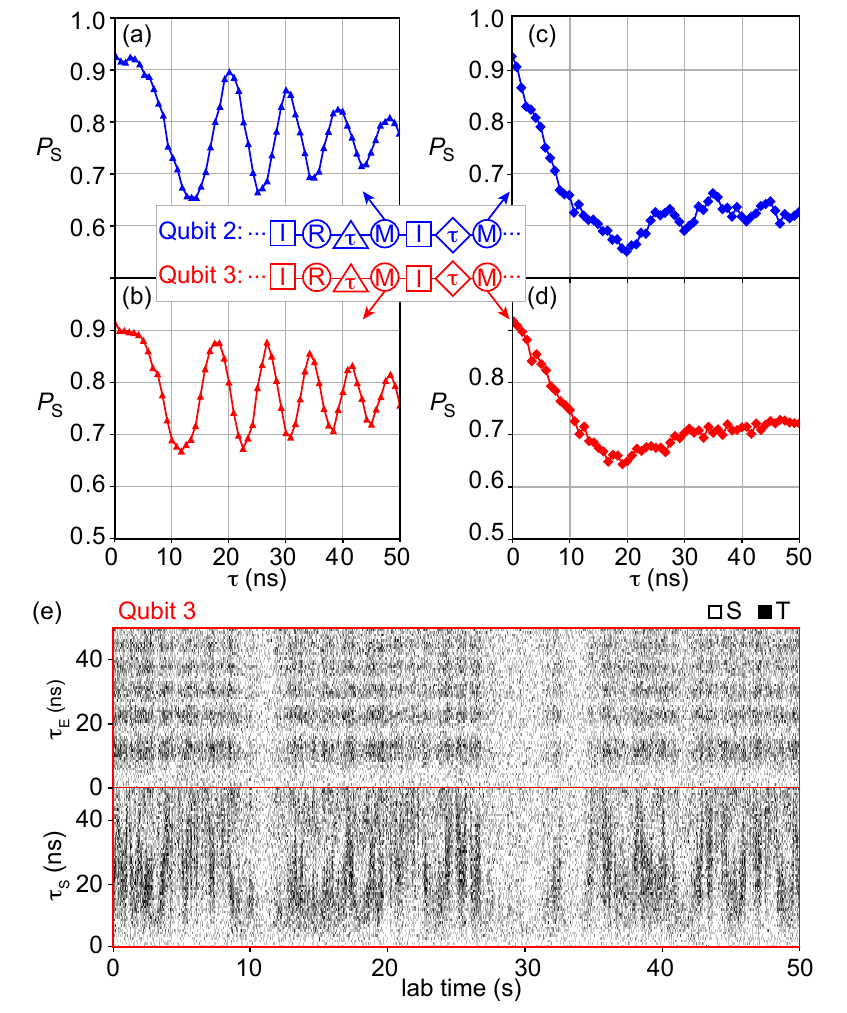}
	\caption{Interleaved qubit operations applied to Q$_2$ and Q$_3$. 
Simultaneous qubit cycles are symbolically represented as in Fig.~\ref{fig2}a. 
(a)-(d) $P_\mathrm{S}(\tau)$ when repeatedly applying an alternating sequence of dephasing and exchange-rotation cycles to both qubits. The value of $\tau$ is changed after each alternation. 
Measurement outcomes following the exchange pulse [dephasing pulse] are represented in panels (a) and (b) [panels (c) and (d)]. 
(e) Non-averaged representation of all single-shot outcomes associated with panel (b) and (d), plotted as a function of $\tau$ for the exchange pulse ($\tau_\mathrm{E}$) or dephasing pulse ($\tau_\mathrm{S}$) and laboratory time. Each singlet outcome for Q$_3$ is represented by one pixel. 
Periods of low Overhauser gradients correlate with periods of low-visibility exchange rotations. 
}
	\label{fig3}
\end{figure}

\subsection{Interleaved operations}

Insight into the finite visibility of exchange oscillations is obtained by alternating qubit exchange cycles and dephasing cycles, as shown in Fig.~\ref{fig3} for two of the qubits. In each alternation cycle, the first measurement bursts yields information about exchange rotations, whereas the second burst yields information about the prevailing Overhauser gradient. For convenience, the pulse duration for the exchange pulse is chosen the same as for the dephasing pulse (parameter $\tau$ in the symbolic qubit cycle in the inset of Fig.~\ref{fig3}), and repeatedly stepped from 0 to 50 ns in 60 steps. 

Plotting the fraction of singlet outcomes from the first measurement burst yields again conventional exchange oscillations for both qubits (Fig.~\ref{fig3}a,b), whereas the second measurement burst yield dephasing data (Fig.~\ref{fig3}c,d). However, if we inspect the underlying single-shot data, as done for Q$_3$ in Fig.~\ref{fig3}e, we observe a temporal correlation between periods of low visibility of exchange oscillations and low Overhauser gradients. We speculate that this indicates a failure of adiabatic conversions between (0,2) singlet states and (1,1) nuclear eigenstates ($\ket{\uparrow\downarrow}$ or $\ket{\downarrow\uparrow}$) in periods of low Overhauser gradients, which could be mitigated by lowering the detuning ramp rate before and after the exchange pulse (cf. Fig.~\ref{fig2}a). 
A similar visibility reduction for exchange oscillations can occur during periods of large Overhauser gradients, via a mechanism that reduces the triplet lifetime in the measurement configuration of the qubit's DQD~\cite{Barthel2012}. 

More generally, knowledge of slowly fluctuating parameters in a multi-qubit processor, obtained from different types of interlaced diagnostic qubit operation cycles, may prove helpful for near-term applications of small and noisy quantum circuits. 

\begin{figure}
	\includegraphics [scale=1] {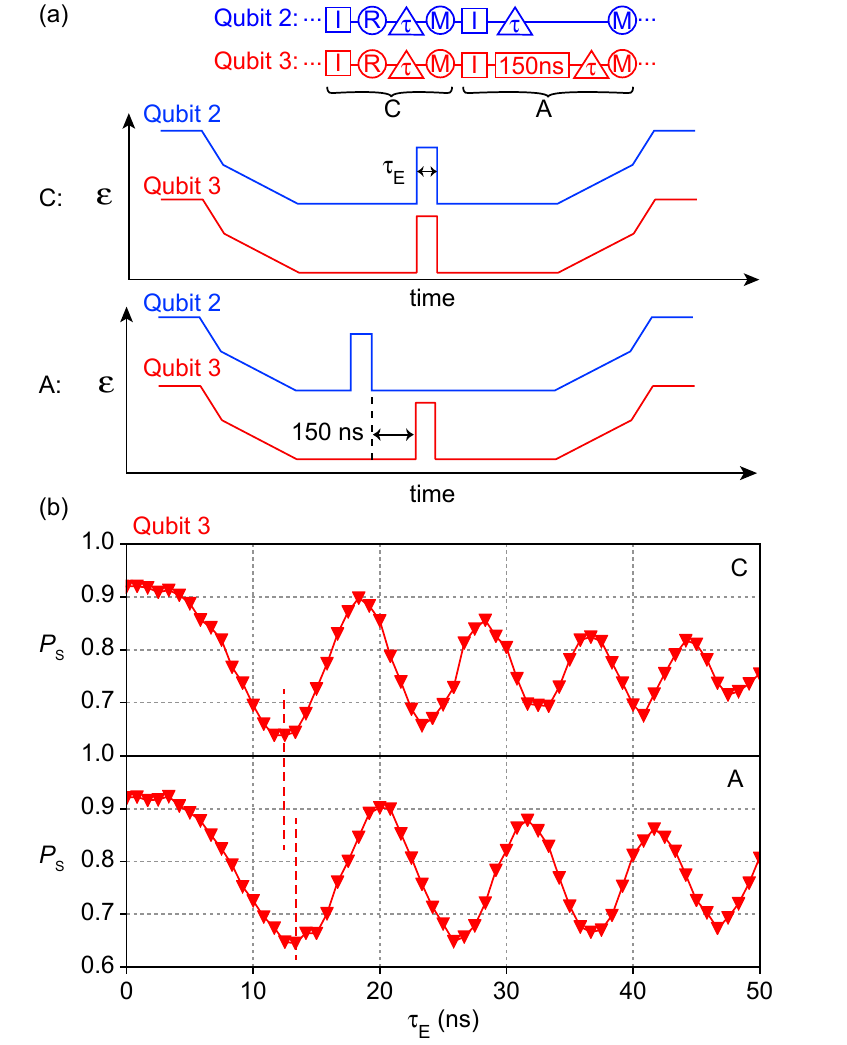}
	\caption[]{
\add{
Crosstalk between qubits. (a) Qubit cycles alternate between protocol C (exchange pulses applied concurrently to Q$_2$ and Q$_3$) and A (pulses applied asynchronously due to a 150-ns delay).  
(b) Q$_3$ exchange oscillations in the presence (C) and absence (A) of  concurrent exchange pulses applied to Q$_2$, indicating that a $\pi$-rotation is obtained 0.9 ns earlier due to crosstalk from Q$_2$ (dashed lines).  
}
	}
	\label{fig4new}
\end{figure}

\add{
\subsection{Crosstalk between qubits}
\label{subsecCrosstalk}	
Fine tuning and optimization of simultaneous qubit operations may require to detect and account for capacitive crosstalk within qubit arrays. We test for qubit-qubit cross talk caused by simultaneous exchange pulses by modifying the interleaved operations of Fig.~\ref{fig3} such that the exchange coupling strength within one qubit (Q$_3$) can be measured in the presence and absence of concurrent exchange pulses applied to its neighbor qubit (Q$_2$). 
Figure~\ref{fig4new}a shows the specific waveforms used for the measurements in ~\ref{fig4new}b, with interaction times $\tau_\mathrm{E}$=0...50~ns varied equally for both qubits in the presence and absence of a 150-ns asynchronization time.   

We observe that the simultaneous exchange pulse applied to Q$_2$ causes a small increase of the exchange coupling within Q$_3$ (which is most obvious in Fig.~\ref{fig4new}b by comparing 6$\pi$ rotations). This cross talk was taken into account (by tuning individual pulse parameters $\varepsilon_{1,2,3,4}$) when demonstrating identical exchange speed for all qubits in Fig.~\ref{fig2}.
}

\subsection{Coupling to the central multielectron dot}

\begin{figure*}
	\includegraphics[scale=1]{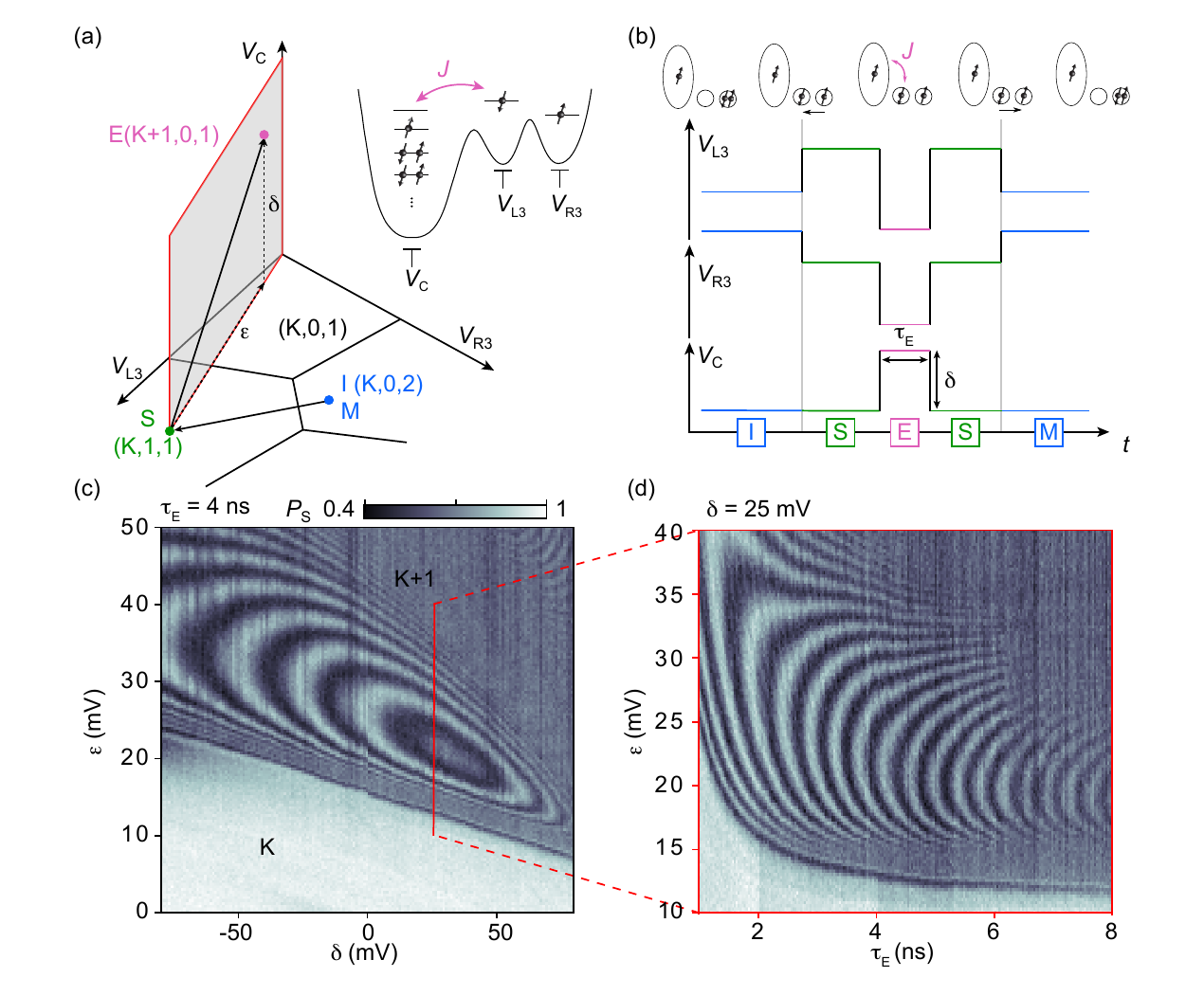}
	\caption{Coherent spin exchange with the multielectron coupler. 
(a) Control voltage trajectory that prepares charge state (K,1,1) with qubit 3 in its spin singlet state (S), before applying simultaneously a detuning pulse ($\varepsilon,\delta$) for duraction $\tau_\mathrm{E}$. For sufficiently large $\varepsilon$ and $\delta$, spin exchange processes (E) induced by the (K+1,0,1) state are expected to reduce the probability of detecting qubit 3 in a singlet state after pulsing via S and Pauli blockade back to M for measurement. 
Inset: Level structure of the qubit-multielectron system for odd occupation number of the coupler, appropriate for results obtained in panels (c,d).
(b) Waveforms $V_{\mathrm{R}3}(t)$, $V_{\mathrm{L}3}(t)$ and $V_\mathrm{M}(t)$ used to implement one manipulation cycle. 
During the exchange pulse (E), $V_\mathrm{C}$ is pulsed positive relative to $V_{\mathrm{L}3}$ and $V_{\mathrm{R}3}$ to attract the left qubit electron to the multielectron dot. 
(c) Fraction of singlet outcomes as a function of $\varepsilon$ and $\delta$ for fixed pulse duration. The diagonal threshold for observing oscillations in $P_\mathrm{S}$ corresponds to the coupler's groundstate transition from odd occupation number (K) to even (K+1). 
A large sweet spot emerges near $\delta$=30 mV that is further characterized in (d) and \ref{figS3}. 
(d) Line cut from (c) for different choices of $\tau_\mathrm{E}$ and fixed detuning $\delta$. 
For all pixels in (c,d), the cycle in (b) is repeated many times to obtain the fraction of singlet measurements. 
}
	\label{fig5}
\end{figure*}

So far we described single qubit operations. To investigate the possibility of two-qubit operations, we now show coherent spin exchange between qubit Q$_3$ and the central multielectron dot (MED). 
\add{Additional spin-exchange data for Q$_2$ and Q$_3$ is provided in the supplement. }

Previous studies in linear arrays of quantum dots demonstrated coherent spin exchange between two singlet-triplet qubits across a MED~\cite{Malinowski2018}. The discrete states of the MED played a central role in mediating the spin-spin coupling, but also gave rise to a variety of electron correlation and spin effects --- including negative exchange couplings~\cite{Martins2017} --- if only one qubit is coupled with the MED. 
In that case, we generally expect that a temporary wavefunction overlap between a qubit spin and a MED with odd occupation results in spin dynamics that changes the state of the qubit spin, which can readily be detected as a suppression of $P_\mathrm{S}$ if the qubit spin was initially part of a singlet pair within the DQD~\cite{MalinowskiPRX2018}. 
(Unlike odd occupation, the ground state of an even occupied MED can be spinless, and Heisenberg spin exchange with a qubit spin may be absent.)

To search for coherent spin exchange with the MED, we therefore pulse one of the qubits (Q$_3$) and the MED while keeping the other three DQDs in Coulomb blockade. 
This allows us to view the pulsed part of the device as a triple quantum dot controlled by $V_{\mathrm{L}3}$, $V_{\mathrm{R}3}$ and $V_\mathrm{C}$ with associated stability diagram as sketched in Fig.~\ref{fig5}a. 
Here, the charge state of this triple dot is indicated by $(m,l,r)$, where $m$ represents the charge state of the multielectron dot while $l$ and $r$ represent the charge state of the left and right well of the DQD. 
  
After finding a suitable (but unknown) configuration of the multielectron dot ($m$=K\add{, see supplement}), 
we demonstrate coherent exchange coupling with the mediator by repeating the following cycle and determining the fraction of singlet outcomes (refer to Fig.~\ref{fig5}b): 
A singlet pair is prepared in the right well and isolated from the reservoirs by Coulomb blockade (I), by keeping the system within (K,0,2). 
The interwell charge transition is then traversed to separate the two entangled electrons from each other (S), by pulsing to (K,1,1) for the duration of one clock cycle of the waveform generator (0.83~ns). 
This operation effectively turns off the intra-DQD exchange interaction while preserving the initialized singlet. 
Then, an exchange pulse  (E) of duration $\tau_\mathrm{E}$ is applied to $V_{\mathrm{L}3}$, $V_{\mathrm{R}3}$ and $V_\mathrm{C}$, with an amplitude parameterized by detuning $\varepsilon$ and $\delta$. Here, $\varepsilon =1/\sqrt{3}\cdot (V_{\mathrm{R}3}-V_\mathrm{R3}^\mathrm{o})+1/\sqrt{3}\cdot(V_{\mathrm{L}3}-V_\mathrm{L3}^\mathrm{o})$ and $\delta = 1/\sqrt{3}\cdot (V_\mathrm{C}-V_\mathrm{C}^\mathrm{o})$, i.e. for sufficiently large $\varepsilon$ and $\delta$ the triple dot transitions to another charge state, (K+1,0,1), thereby maximizing the wave function overlap between the left qubit spin and the spin of the multielectron dot (Fig.~\ref{fig5}a, inset). 
(Here, $V_j^\mathrm{o}$ are DC operating voltages associated with the three gate electrodes.) 
The qubit spin is returned to the left well, by pulsing to the separation point in (K,1,1) for one clock cycle, and read out by spin-to-charge conversion by pulsing towards (K,0,2). Any change of the qubit spin during the exchange pulse, coherent or incoherent, is expected to show up as a reduction of $P_\mathrm{S}$ during the measurement burst (M). 

A plot of $P_\mathrm{S}(\varepsilon,\delta)$ for a fixed exchange pulse duration of 4~ns clearly shows a reduction above a critical pulse-amplitude threshold (sloped feature in Fig.~\ref{fig5}c), which we associate with the charge transition between (K,1,1) and (K+1,0,1). Above that threshold, $P_\mathrm{S}$ oscillates as a function of $\varepsilon$ and $\delta$, which we regard as evidence for coherent spin exchange processes from the MED~\cite{MalinowskiPRX2018}. 

The ring-like features in $P_\mathrm{S}(\varepsilon,\delta, \tau_\mathrm{E}$=constant) can be viewed as contours of constant exchange coupling strength, revealing a non-monotonic dependence of the exchange coupling on detuning and a prominent maximum. This is also evident if $\tau_\mathrm{E}$ is varied for fixed $\delta$ (Fig.~\ref{fig5}d). A qualitative comparison between this behaviour with a systematic study of a similar multielectron dot~\cite{MalinowskiPRX2018} suggests that the MED occupation K is odd with an effective ground state spin of 1/2. 
In this interpretation, the emergence of a sweet spot and a sign reversal of the exchange coupling strength for larger detuning amplitudes arises from a competition of electron correlations on the multielectron dot and its relatively small level spacing~\cite{Martins2017}.

\section{Discussion}
In this work, using frequency-multiplexed high-frequency reflectometry sensors and 8-dimensional gate-voltage pulses, we demonstrate for the first time the simultaneous coherent manipulation and readout of four spin qubits. 

Arranged in a 2$\times$2 two-dimensional geometry and implemented as singlet-triplet qubits in GaAs, this allows us to sense temporal Overhauser fluctuations at four different sample locations. 
Our data does not show clear correlations between the four local GaAs nuclear environments. This supports the picture that statistical nuclear polarizations are local~\cite{Taylor2007} and is consistent with nuclear spin diffusion occurring predominantly perpendicular to the plane of the 2DEG, into the heterostructure~\cite{Petta2008}. 
However, we did not intentionally induce large nuclear polarizations, as previously done in single-qubit experiments~\cite{Bluhm2010}. 
Our demonstration of spatio-temporal quantum sensing exemplifies the utility of qubit arrays beyond quantum computation, and can be refined in straightforward ways~\cite{Luke2016}. 

In addition to simultaneous exchange rotations within the array (with all qubits tuned to the same rotation speed), we also explored the diagnostic value of interleaving different qubit operations. Here, the alternation of Overhauser rotations and exchange rotations revealed potential errors in qubit initialization, providing additional insights for optimizing qubit fidelities that may supplement existing techniques, such as closed-loop control~\cite{Cerfontaine2020self,Cerfontaine2020}. 
\add{A modification of interleaved operations then allowed us to detect a small change of exchange strengths in the presence of exchange pulses applied to other qubits, which we attribute to a small crosstalk from distant gate electrodes. We expect that automatic compensation of crosstalk will become even more important for controlling larger or denser qubit arrays at high fidelity. }


Lastly, we demonstrated coherent spin exchange between one of the qubits and the central multielectron dot. The observed exchange profile is qualitatively in agreement with previous studies in linear arrays~\cite{Martins2017,MalinowskiPRX2018} for an odd-occupied multielectron dot with an effective spin-1/2 ground state. 
\add{However, additional data (supplement) suggests a prevalence of high-spin ground states that possibly arises from an enhancement of interaction effects by the large size of the multielectron dot, which in future experiments may affect the search for zero-spin ground states suitable for coherent two-qubit coupling~\cite{Malinowski2018}. }

Overall, our results suggest that large quantum dots may prove useful as tunable inter-qubit couplers to realize two-dimensional qubit networks. However, the large number of gate voltages that need to be tuned and synchronized is currently challenging for our manual tune-up procedures, and motivates the development of super-human automation~\cite{Moon2020}.  

\section{Author contributions}
S.F., G.C.G. and M.J.M. grew the heterostructure. F.F. fabricated the device. F.F., A.C., performed the experiments. F.F., A.C., F.K. analyzed data and prepared the manuscript.

\section{Acknowledgements}
We thank Fabio Ansaloni for technical help and acknowledge support from the Innovation Fund Denmark and the Independent Research Fund Denmark. 
A.C. acknowledges support from the EPSRC Doctoral Prize Fellowship. 


\section{Methods}

\subsection{Material and device fabrication}
The device was fabricated on a GaAs/AlGaAs heterostructure with a 2DEG 57 nm below the surface with carrier density $n = 2.5 $ x $10^{15}$ m$^{-2}$ and mobility $\mu = 230$ m$^2/$Vs . The qubit array was realised using standard electron-beam lithography to pattern Ti/Au metallic gates, after the deposition of a 10-nm thick layer of HfO$_2$. This insulating layer allows the application of both positive and negative gate voltages and obviates the need of bias cooling~\cite{Martins2016}. 
For Figures 1-4 the accumulation gate was held at +50 mV, while it was set to +75 mV for Figure~\ref{fig5}. 

\subsection{Synchronization of 8-dimensional voltage pulses}
The four qubits are simultaneously controlled by application of 8-dimensional gate-voltage pulses. This required sub-nanosecond synchronization of two four-channel arbitrary waveform generators (Tektronix 5014C), which we achieved by triggering both generators with voltage pulses applied via identical cables. In addition, each instrument was synchronized to a 10-MHz frequency standard (SRS FS725). 

\subsection{Four-qubit single-shot readout}
To acquire and analyze the demodulated voltage of all four sensor quantum dots during each waveform cycle, we extended previous approaches for frequency multiplexing~\cite{Laird2010} (see supplementary Figure~\ref{figS1} for details) and state identification~\cite{Malinowski2018}. 
Specifically, during each operation cycle, the four reflectometry tones are temporarily switched on before and after the separation pulse. This allows the acquisition of one reference measurement ($V_\mathrm{RF}^\mathrm{R}$) after the singlet initialization, and one actual measurement ($V_\mathrm{RF}^\mathrm{M}$) after the qubit operation. 
To assign single-shot outcomes, the demodulated voltage of each reference burst is subtracted from the demodulated voltage of the measurement burst. The outcomes are then collected into a histogram and fitted with a double Gaussian function. 
The midpoint between the two Gaussians serves as a voltage threshold, i.e. we assign all outcomes on the singlet (triplet) side to 1 (0). 
This identification of qubit states is similar to single-qubit readout in earlier experiments~\cite{Barthel2010}, with the additional improvement that the subtraction prior to histogramming makes our method insensitive to small and slow drifts of the sensor operating points, allowing the execution of many qubit cycles over long periods of time. 

\add{Similar to previous work~\cite{Laird2010}, we find that the depths of the readout resonances in Figure~\ref{fig2}b depend on the tuning (predominantly conductance) of the sensor dots. To improve single-shot readout of the experiments reported in this work, each sensor tuning had been optimized relative to the configuration of Figure~\ref{fig2}b. }

\clearpage

\renewcommand{\thefigure}{S\arabic{figure}}
\setcounter{figure}{0}
\onecolumngrid
\appendix

\section{Supplementary Information for "Simultaneous operations in a two-dimensional array of singlet-triplet qubits"}	


\subsection{Demodulation of four carrier frequencies}	
\begin{figure*}[h]
	\centering
	\includegraphics [width=1.0\linewidth] {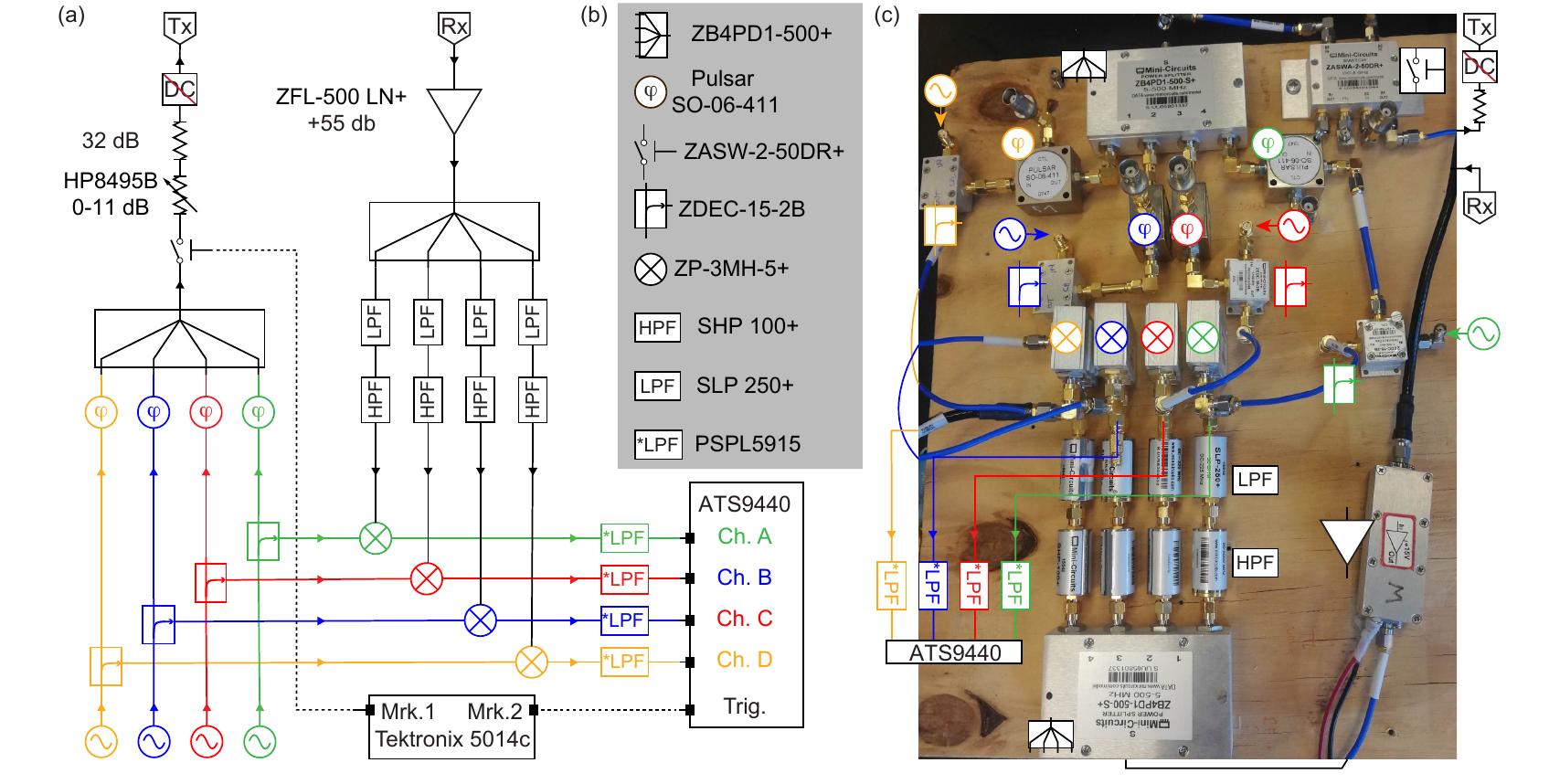}
	\caption[]{Four-channel demodulation circuit. (a) Schematic of the circuit layout. Colored lines indicate paths associated with different carrier frequencies, while solid lines indicate paths in which all four rf tones propagate simultaneously. The color code follows the one used in the main text. 
	(b) Legend of specific components used in this work. 
	(c) Photograph of the circuit layout. 
	The transmitter and receiver port of the cryostat is denoted by Tx and Rx, respectively. }
	\label{figS1}
\end{figure*}

To perform simultaneous qubit readout we extended to four channels a frequency multiplexing setup similar to that in Ref.~\onlinecite{Laird2010}. 
Our setup is shown in Figure~\ref{figS1} as a circuit schematic and as a photograph of the demodulation circuit. 

Four carrier frequencies are generated using independent RF sources (SRS386). Each frequency is passed through a directional coupler that separates the probe tone from the reference tone. Each probe tone is then passed through a voltage controlled phased shifter, before being added with the other probe tones using a four-port power splitter. The resulting signal now contains all four carrier frequencies and is passed through an high-frequency switch, followed by a series of attenuators and filters before being injected into the transmission port (Tx) of the cryostat. Inside the cryostat (not shown), the four carriers are further attenuated before reaching a directional coupler that applies a fraction of the signal to four LC resonators located on the PCB sample holder~\cite{Qdevil}.  Each resonator is wirebonded to an ohmic contact of one of the four sensor quantum dots. Since each resonator is excited only by one of the carrier frequencies (whichever one is resonant with the LC circuit), this provides qubit addressability. 

After being reflected by the sample the four carriers are directed by the directional coupler to a cryogenic amplifier before exiting the cryostat at its receiver port (Rx). The signal is then further amplified at room temperature before being split into four equal components by a four-port power splitter. Finally, each component is mixed with one of the reference tones. The demodulated output of each mixer is then low-pass filtered and recorded by a fast digitizer card (Alazar ATS9440).

To implement the reference and measurement bursts described in the method section of the main text, we drive the high-frequency switch located before the Tx port with a marker channel of the arbitrary waveform generator (dashed black line in Figure~\ref{figS1}a). 
A second marker is connected to the trigger port of the Alazar card to mark the beginning of a  acquisition. To account for an inevitable delay between the triggers and the instant the signal is received from the demodulation circuit, caused by propagation delays associated with cables and cryostat wiring, we adjust the post-trigger delay of the Alazar card such that its sampling starts after the trigger is received (in our case 500 ns). 

\add{
\subsection{Additional data for spin exchange with the multielectron dot}	
In Figure~\ref{figS2}a we show the demodulated voltage measured from S$_3$ (after the subtraction of a background plane) as a function of $V_\mathrm{C}$ and $V_{\mathrm{L}3}$. The resulting charge stability diagram shows horizontal and vertical transitions associated with charging events of the left dot of qubit 3 and the multielectron dot, respectively. White dashed lines highlight inter-dot charge transitions that can be traversed (along suitable detuning directions $\varepsilon$) to test for coherent coupling with the multielectron dot. 
These data were taken column by column, to allow the multielectron dot sufficient time to exchange electrons with one of the reservoirs while $V_\mathrm{C}$ is slowly changed, likely via leakage through the Q$_1$ or Q$_4$ double dot. 

\begin{figure*}[h]
	\centering
	\includegraphics [width=0.9\linewidth] {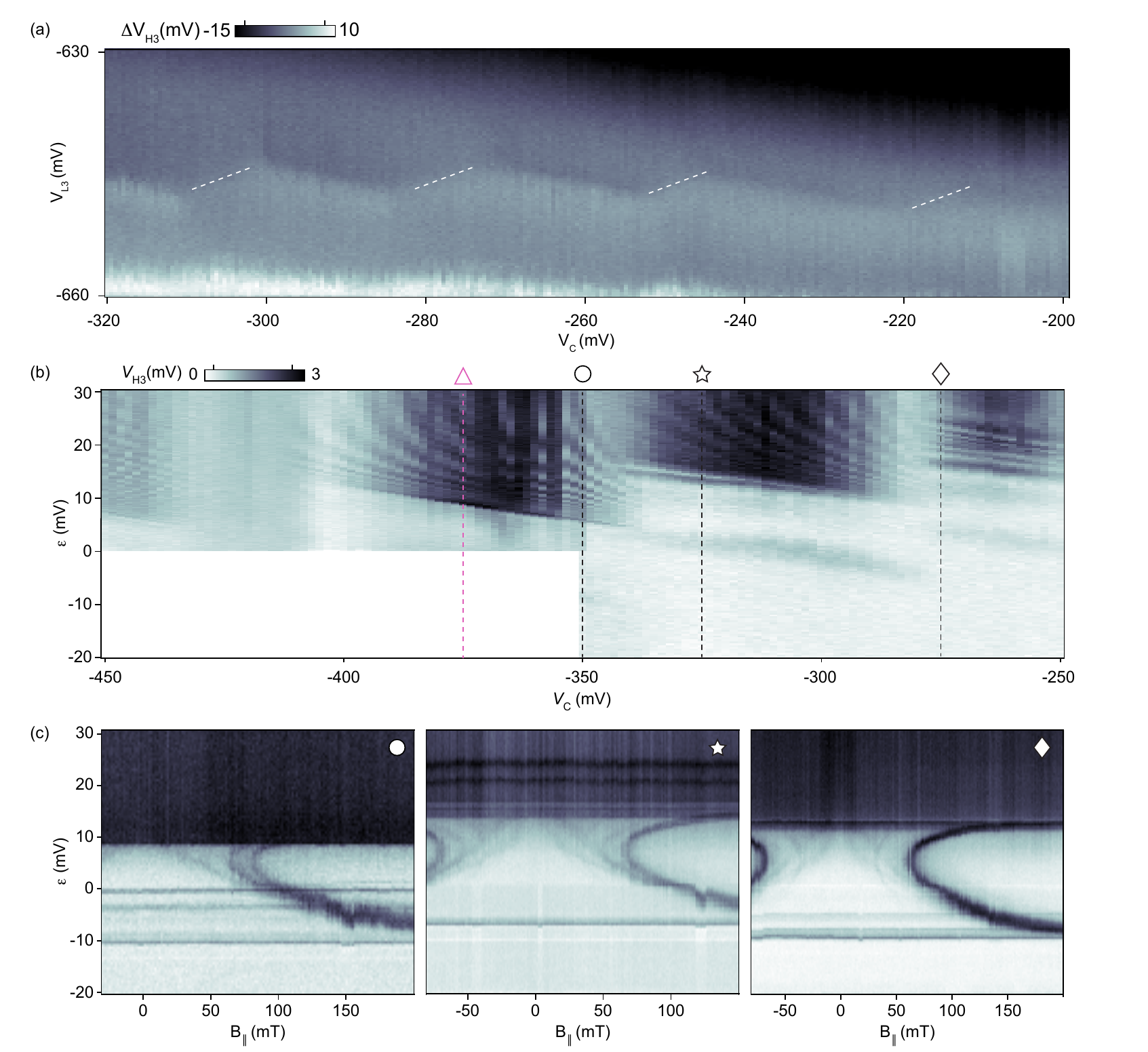}
	\caption[]{
\add{
Coherent spin exchange and leakage spectroscopy of the multielectron dot for several charge occupations. (a) Charge stability diagram of the multielectron dot (MED) as a function of $V_\mathrm{C}$ and $V_{\mathrm{L}3}$ in the absence of pulses. Dashed lines indicate inter-dot transitions between the left dot of qubit 3 and the MED. 
(b) Demodulated voltage $V_{\mathrm{H}3}$ as a function of detuning parameter $\varepsilon$ and $V_\mathrm{C}$ for $\tau_{\mathrm{E}}$=4~ns, indicating coherent exchange oscillations in several consecutive charge occupations of the MED. 
The triangular marker indicates $V_\mathrm{C}$ for the spin exchange presented in Figures~\ref{fig5}, \ref{figS3} and \ref{figS4}b. 
(c) Leakage spectroscopy data (using a fixed interaction time $\tau_{\mathrm{E}}$=150~ns and $V_\mathrm{C}$ values marked in panel b) showing $V_{\mathrm{H}3}$ as a function of in-plane magnetic field B$_\parallel$ and detuning $\varepsilon$ for three consecutive charge occupations of the multielectron electron dot. 
In (b) and (c), enhancements of $V_{\mathrm{H}3}$ correspond to a suppression of $P_\mathrm{S}$, i.e. spin exchange or leakage. 
}
	}
	\label{figS2} 
\end{figure*}

In Figure~\ref{figS2}b we use the same exchange sequence used for Figure~\ref{fig5} and measure the demodulated voltage of S$_3$ as a function of the detuning parameter $\varepsilon$ and the MED plunger gate $V_\mathrm{C}$. As function of $V_\mathrm{C}$ the number of electrons in the MED changes by one approximately every 40-50 mV. 
Within one charge state, an increase in $V_\mathrm{C}$ has the same effect as an increase of $\delta$ in Figure~\ref{fig5}d. We can clearly observe coherent exchange oscillations in several consecutive charge occupations. 
Since coherent oscillations indicate the presence of an unpaired spin in the MED, this behavior is inconsistent with a single-particle picture of the multielectron dot in which the ground-state spin alternates between S=0 and S=1/2. Deviations from this simple non-interacting picture have previously been observed by Malinowski~\emph{et al.} using a single-triplet qubit coupled to a multielectron dot~\cite{MalinowskiPRX2018}. We speculate that the ground state of the MED alternates between S=1/2 and S=1 in this region.   

The presence of non-zero spin configurations in all three consecutive occupations is further confirmed by leakage spectroscopy measurements, see Figure~\ref{figS2}c. Leakage spectroscopy data are obtained by pulsing qubit 3 and the MED with the same waveforms used in the main text (see Fig.~\ref{fig5}b) but with longer interaction time (fixed at $\tau_{\mathrm{E}}$=150~ns, sufficiently long to detect incoherent spin mixing between the MED and the qubit electron). 
Markers indicate the choice of $V_\mathrm{C}$ for each measurement, whereas the triangle indicates the value of $V_\mathrm{C}$ for the data presented in Fig.~\ref{fig5}c-d. 

As a function of detuning $\varepsilon$ and in-plane magnetic field B$_{\parallel}$, we observe in all three cases a U-like leakage feature. Similar features arise in three-electron triple dots (in Ref.~\onlinecite{medford2013} the feature arises from the crossing of two states with different spin projections, denoted as $\ket{\uparrow}{S}$ with $\ket{\uparrow\uparrow\uparrow}$) and have also been observed by coupling a singlet-triplet qubit to a spin-1/2 and spin-1 multielectron dot~\cite{Martins2017,MalinowskiPRX2018}. \\


Figure~\ref{figS3} shows the effect of applied magnetic field on the exchange profile of Figure~\ref{fig5}d. 
Regarding contours of equal $P_\mathrm{S}$ as points of constant exchange coupling strength, we observe that 
 in-plane (B$_{\parallel}$) and out-of-plane (B$_{\perp}$) magnetic fields have no strong effect on the exchange strength. The weak dependence on B$_{\parallel}$ is consistent with similar observations by Martins~\emph{et. al.} for a linear array with somewhat smaller multielectron dot, although they found a stronger dependence on B$_{\perp}$ that was vaguely attributed to orbital magnetic effects~\cite{Martins2017}. 
 
Orbital effects can potentially be studied by changing the size or shape of the multielectron dot by tuning the four large depletion gates ("backbone gates") that are surrounding the elliptical area indicated in Figure~\ref{fig1}a. We have not systematically studied this capability, but note that for experiments in Fig.~\ref{figS2},\ref{figS3},\ref{figS4} and \ref{fig5} the depletion voltages applied to the upper two backbone gates (-0.70~V for Q$_1$ and Q$_4$) were more negative relative to the lower two backbone gates (-0.54 V and -0.56 V for Q$_2$ and Q$_3$). 
In this regime, we do not have direct experimental evidence that the multielectron dot extends all the way underneath the ellipse indicated in Figure~\ref{fig1}a. (In the regime where tunnel barriers associated with upper and lower qubits are low, and where upper backbone voltages are comparable to lower backbone voltages, we did verify Coulomb blockade of the fully-extended quantum dot by measuring transport between ohmics located on the upper and lower sides of the chip.) 

\begin{figure*}[h]
	\centering
	\includegraphics [width=0.8\linewidth] {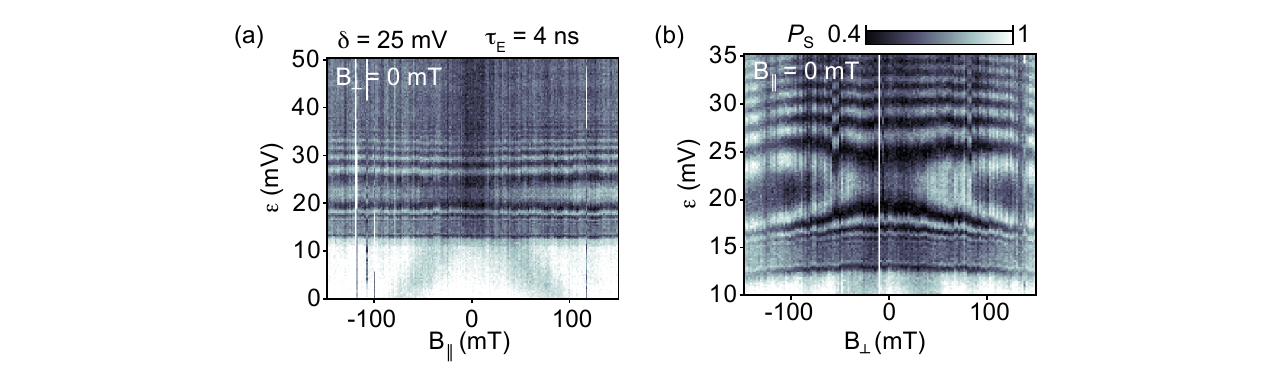}
	\caption[]{
\add{
Dependence of the exchange pattern in Figure ~\ref{fig5} on applied magnetic field, for a fixed amplitude $\delta$=25~mV and interaction time $\tau_\mathrm{E}$=4~ns. 
(a) Fraction of singlet outcomes as a function of $\varepsilon$ and B$_\parallel$ for B$_\perp$=0. 
(b) Fraction of singlet outcomes as a function of $\varepsilon$ and B$_\perp$ for B$_\parallel$=0. 
}
	}
	\label{figS3}
\end{figure*}

\begin{figure*}[h]
	\centering
	\includegraphics [width=0.5\linewidth] {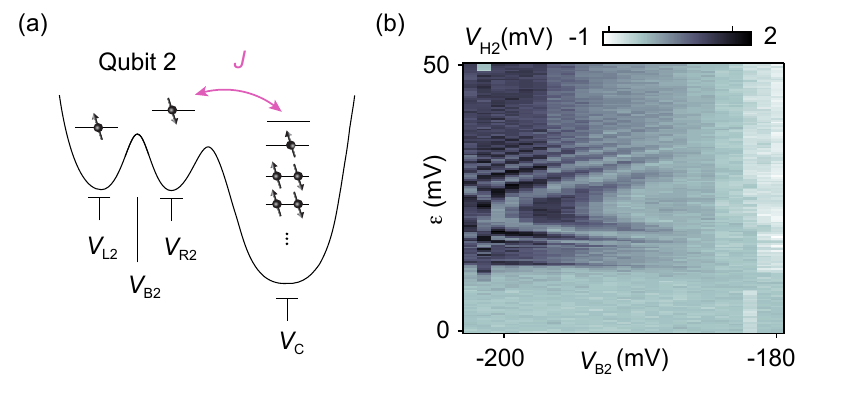}
	\caption[]{
\add{
Coherent spin exchange between Q$_2$ and the multielectron dot. (a) Level structure of the qubit-multielectron system similar to the one presented for Q$_3$ in Figure~\ref{fig5}.  $V_{\mathrm{B2}}$ is the voltage applied to the interdot barrier gate of Q$_2$. (b) Demodulated voltage $V_{\mathrm{H}2}$ as a function of detuning $\varepsilon$ and $V_{\mathrm{B2}}$. 
For this tuning of sensor 2, an enhanced value of $V_{\mathrm{H}2}$ indicates a suppression of $P_\mathrm{S}$. 
}
	}
	\label{figS4}
\end{figure*}


The pulse sequence described in Figure~\ref{fig5} can also be applied to other qubits. For example, by pulsing $V_{\mathrm{R}2}$ and $V_{\mathrm{L}2}$ instead of $V_{\mathrm{L}3}$ and $V_{\mathrm{R}3}$ we can view the pulsed part of the device as a triple quantum dot comprised of Q$_2$ and the multielectron dot (Fig.~\ref{figS4}a), while  Q$_3$ remains in Coulomb blockade. 
The resulting coherent exchange oscillations from Q$_2$ (detected by measuring $V_{\mathrm{H}2}$) are presented as columns in Figure~\ref{figS4}b, for different voltages applied to the DQD barrier gate ($V_{\mathrm{B}2}$). The dependence of exchange oscillations on $V_{\mathrm{B}2}$ suggests that the effective detuning of the multielectron dot and the resulting exchange coupling strength are affected somewhat by capacitive cross coupling from the barrier gate. 
During this experiment, all other tuning voltages are identical to those used for the Q$_3$ spin-exchange experiments (corresponding to the triangular marker in Figure~\ref{figS2}b), indicating that multi-qubit coupling to the multielectron dot is possible. 
However, we have not yet demonstrated qubit-qubit coupling mediated by the multielectron dot (ideally the coupling of any qubit to any other qubit), but expect that more efficient tuning techniques (for example based on automation) will allow this in future. 

}

\subsection{T$_2^{*}$ fits and exchange quality factors}	

Fits for the time averaged data of Figure~\ref{fig2}c in the main text are adapted from Ref.~\onlinecite{Petta2005} using an exponentially damped sine function: 
\begin{equation}
P_{\mathrm{S}}(\tau) = C_1 + C_2 \sin(2\pi f\tau + C_3)e^{-(\frac{\tau}{T_2^{*}})^2} 
\end{equation}
with $C_{1,2,3}$, $f$ and  $T_2^{*}$ as free parameters. Here, $T_2^{*}$ quantifies the inhomogenous dephasing time due to hyperfine interactions with nuclear spins. 

The fits for the exchange oscillations in Figure~\ref{fig2}d are obtained using the function: 
\begin{equation}
P_{\mathrm{S}}(\tau) = C_1 + C_2 \cos(2\pi J\tau + C_3)e^{-(\frac{\tau}{T_\mathrm{el}^{*}})^2} 
\label{EqFitJ}
\end{equation}
with $C_{1,2,3}$, $J$ and  $T_\mathrm{el}^{*}$ as free parameters. 
\add{
Here, $T_\mathrm{el}^{*}$ is sensitive to decoherence arising from noise in the exchange coupling strength $J$, and likely has contributions from effective electrical noise. 
Quality factors are obtained by  $Q=J \cdot T_\mathrm{el}^{*}$ following Ref.~\onlinecite{Martins2016}. 
Previous work in Ref.~\onlinecite{Malinowski2018} has shown that the high-frequency wiring of the cryostat results in a finite rise time of exchange pulses, which results in deviations from Eq.~\ref{EqFitJ} for short values of $\tau$. In Figure~\ref{fig2}d, we therefore exclude data of the first 15 ns from the fits. }

\end{document}